# Semiconductor Bowtie Nanoantenna from Coupled Colloidal Quantum Dot Molecules

Jiabin Cui[a, b], Somnath Koley[a, b], Yossef E. Panfil[a, b], Adar Levi[a, b], Nir Waiskopf[a, b], Sergei Remennik[b], Meirav Oded[a, b] & Uri Banin*[a, b]

[a]  Dr, J. B. Cui, Dr, S. Koley, Mr, Y. E. Panfil, Mr, A. Levi, Dr, N. Waiskopf, Dr, M. Oded, Prof. U. Banin
Institute of Chemistry
The Hebrew University of Jerusalem
Jerusalem 91904, Israel.
E-mail: uri.banin@mail.huji.ac.il (U.B.)

[b]  Dr, J. B. Cui, Dr, S. Koley, Mr, Y. E. Panfil, Mr, A. Levi, Dr, N. Waiskopf, Dr, S. Remennik, Dr, M. Oded, Prof. U. Banin
The Center for Nanoscience and Nanotechnology
The Hebrew University of Jerusalem
Jerusalem 91904, Israel.

Supporting information for this article is given via a link at the end of the document.

**Abstract:** Top-down fabricated nanoantenna architectures of both metallic and dielectric materials demonstrated powerful functionalities for Raman and fluorescence enhancement with relevance to single molecule sensing, while inducing directionality of chromophore emission with implications for single photon sources. Herein, we synthesize the smallest bowtie nanoantenna by selective tip-to-tip fusion of two tetrahedral colloidal quantum dots (CQDs) forming a dimer. While the tetrahedral monomers emit non-polarized light, the bowtie architecture manifests nanoantenna functionality of enhanced emission polarization along the bowtie axis as predicted theoretically and revealed by single particle spectroscopy. Theory also predicts the formation of an electric-field hotspot at the bowtie epicenter. This is utilized for selective light induced photocatalytic metal growth at that location, unlike growth on the free tips in dark conditions thus demonstrating the bowtie dimer functionality as a photochemical reaction center. Our findings pave a path for additional bottom-up bowtie architectures applicable in optics, sensing and photocatalysis.

## Introduction

Optical nanoantennas are devices that concentrate and direct the electromagnetic radiation to nano- and atomic-length scales facilitating potential applications in microscopy, optical nanocircuitry, single photon sources, sensing, light harvesting and optoelectronics.[1–4] A highly efficient system is the metallic nanoantenna, typically prepared by e-beam lithography, that manifests exceptional near field enhancement enabling ultimate single molecule optical and chemical sensing as well as plasmonic photocatalysis.[6–10] Dielectric nanoparticle architectures, either as arrays or in dimers, also exhibit electric and magnetic field hotspots, due to the hybridization of the electromagnetic modes.[3,10-11] Dielectric Galium Phosphide nanoantenna reported to possess useful functionalities, overcoming the extensive heat genaration and radiative loss present in the high dielecric metal alternatives. This is interesting to realize what functionalities a low dielectric nanoantenna, scaled-down to an order of magnitude, can manifest.[12] In this quest for ultimate electromagnetic field concentration by nanoantennas to efficiently couple visible light photons in and out of nanoscale objects, the bowtie architecture yields the most significant enhancement characteristics surpassing structures prepared from spherical or rod-like geometries.[13],[14] Here, we demonstrate the bottom-up synthesis of a scaled down semiconductor bowtie nanoantenna constructed from colloidal quantum dots (CQDs), utilizing the inherent tip reactivity of the tetrahedral CQD building blocks. Various bowtie morphologies were identified, all mapped to twisting and scissoring modes of the truncated tetrahedral CQD units. The bowtie nanodimer manifested enhanced single particle emission polarization along the dimer axis, unlike the non-polarized emission of the tetrahedral monomers, and in correspondence with theoretical calculations of the geometry related electric field distribution. Theory also predicted an electric field hotspot at the bowtie epicenter, which is utilized as a photochemical reaction center achieving selective light induced metal growth at this region. This study further demonstrates the strength of "Nanocrystals chemistry" that entails the rational adjoining of quantum sized nanocrystals serving as artificial atoms to sculpt novel architectures with desired properties, in which the tunable building blocks and their tailored connectivity yield novel electronic, optical and also photochemical functionalities.
The strength of the nanocrystals chemistry approach is based on the utilization of the highly developed CQDs as building blocks with well-controlled crystal lattice, size and shape that can be assembled to complex architectures via control of their surface chemistry.[15] In particular, fusing the adjacent CQD building blocks has been utilized to form arrays,[16,17] honeycomb superlattices,[18] elongated coupled hetero-structures,[19] and artificial dimer molecules.[20] Such epitaxial fusion of spherical core/shell nanocrystals with wurtzite crystal lattice, yielded homodimer molecules where quantum mechanical tunneling facilitated electron wavefunction hybridization between the two adjacent CQDs. By utilizing the colloidal approach for the controlled adjunction of tetrahedral quantum dots with zinc-blende crystal lattice we achieve the desired bowtie nanoantenna architecture. Unlike the wurtzite CQDs with diverse facets, these tetrahedral CQDs present indistinguishable surface facets with only two possible attachment scenarios: either through the identical facets or through the tips. The attachment of tetrahedral CQDs through the facets was found to be driven by the thermodynamic parameters at the lattice-ligand interface to form quasi-crystal morphologies[21,22] or by dipolar interactions utilizing the interfacial charge promoting oxidation of the surface atoms to form twisted ribbons[23]. Herein, by utilizing the high chemical reactivity of the apexes we achieve kinetically driven highly selective tip attachment of the tetrahedral CQDs, thus forming the bowtie dimers.

## Results and Discussion

The bowtie CQD structures were prepared in solution by first, forming chemically linked dimers on silica spheres serving as a soft template, followed by their fusion to form a continuous lattice (Figure 1a, Figure S1). The model building blocks are CdSe/CdS core/shell CQDs with zinc-blende crystal lattice [24] of tetrahedral shape with slightly truncated edges (Figure 1b, Figure S2-4). Briefly, the tetrahedral CQDs were first linked to the $SiO_2$ nanospheres, preferentially with the base facing the silica surface on account of the contribution of multiple binding sites to the CQD facet (Figure S5). A protective thin silica layer was deposited on the nanospheres to immobilize the CQDs followed by a tetrathiol cross-linker binding at their exposed tip. Upon stoichiometric addition of the secondary tetrahedral CQDs, dimers, linked through the tips, were formed (Fig. 1c, Figure S6-7). Subsequently, selective chemical etching of the $SiO_2$ nanospheres led to release of the CQD dimers (Figure S8). Epitaxial fusion to generate robust bowtie CQD structures was carried out at high temperature along with addition of excess Cd-oleate (Figure 1d). The favorable temperature window for fusion of the zinc-blende CQDs is 180-210 °C, notably narrower than the case of the wurtzite dimers on account of their higher reactivity (Figure S9). Two to three cycles of size selective separation were performed to achieve a fraction with high percentage of dimers (Figure S10).

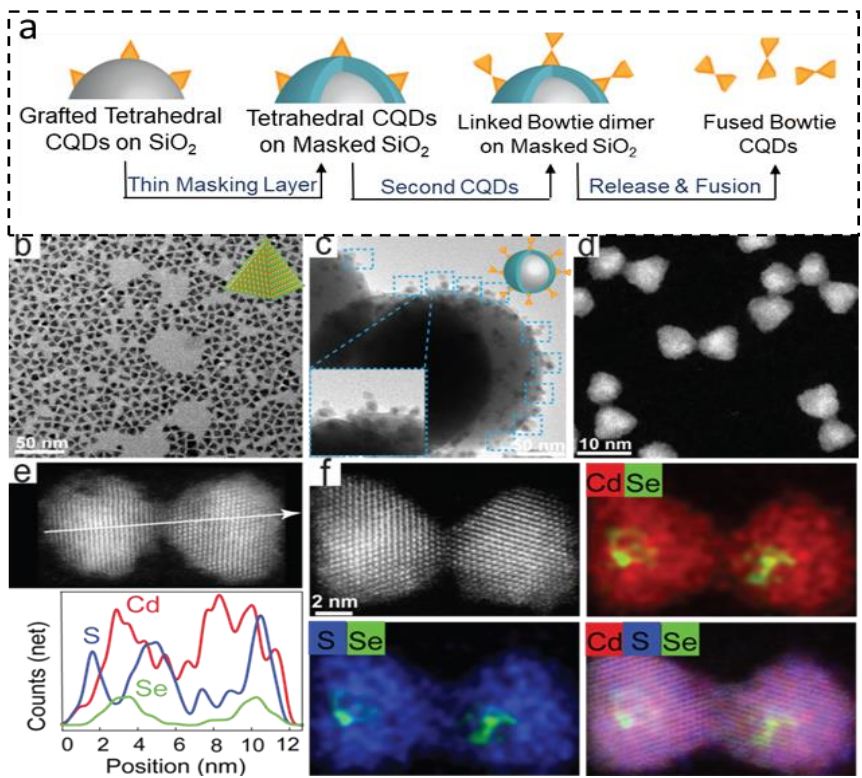

**Figure. 1.** Fabrication of Bowtie CdSe/CdS CQDs. (a). Simplified scheme for the synthesis of the bowtie dimer from tetrahedral CQDs monomers (Detailed in Figure S1 and discussion therein). (b) TEM image of monomer zinc-blende CdSe/CdS core/shell CQDs with the structural model in the inset. (c) TEM image of chemically linked dimer on soft silica template (CdSe/CdS@$SiO_2$) as represented in the cartoon at top right. The blue frames in the selected squares highlights the dimer structures on the surface of the $SiO_2$. (d) STEM image of the fused bowtie CQDs. (e) HAADF-EDS line scan and (f) STEM-EDS elemental mapping of the structurally fused CdSe/CdS bowtie architecture. The core locations are indicated by the localized Se signals.

The bowtie architecture was characterized and established by high-angle annular dark field-scanning TEM (HAADF-STEM) and energy dispersive spectroscopy (EDS) analysis measurements (Figure 1e-f, Figure S10-11). Continuous atomic lattice fringes are observed throughout the dimer structure. The elemental maps in the bowtie demonstrate continuous distribution of cadmium in the entire structure with the selenium and sulfur elements dispersed in the core and shell regions, respectively. This establishes the preservation of the core/shell architecture throughout the bowtie synthesis process.

In order to prove and comprehend the binding relationship and inherent attachment mechanism forming the bowtie architecture, the 3D atomic structure was reconstructed by HAADF-STEM imaging, Fast Fourier Transform (FFT) analysis and a simulated reconstruction model (Figure 2). We first establish the tetrahedral structure of the CQD monomers from the projections manifested by the STEM images. Four main projection planes are resolved for the randomly oriented monomer CQDs on the TEM grid (Fig. 2a-d). All these can be directly mapped to a truncated tetrahedral shaped zinc-blende monomer structure with (111) surface facets as depicted in Fig. 2e (Figure S12, Supplementary Table 1-4 for detailed analysis). Projections (i-iv) shown in frames a-d, correspond to viewing the monomer along different zone axes (ZA): [100], [110], [111] and [11$\bar{2}$], respectively, matching well to the simulated 3D structure in Fig. 2e illustrating the orientations leading to the observed projections.

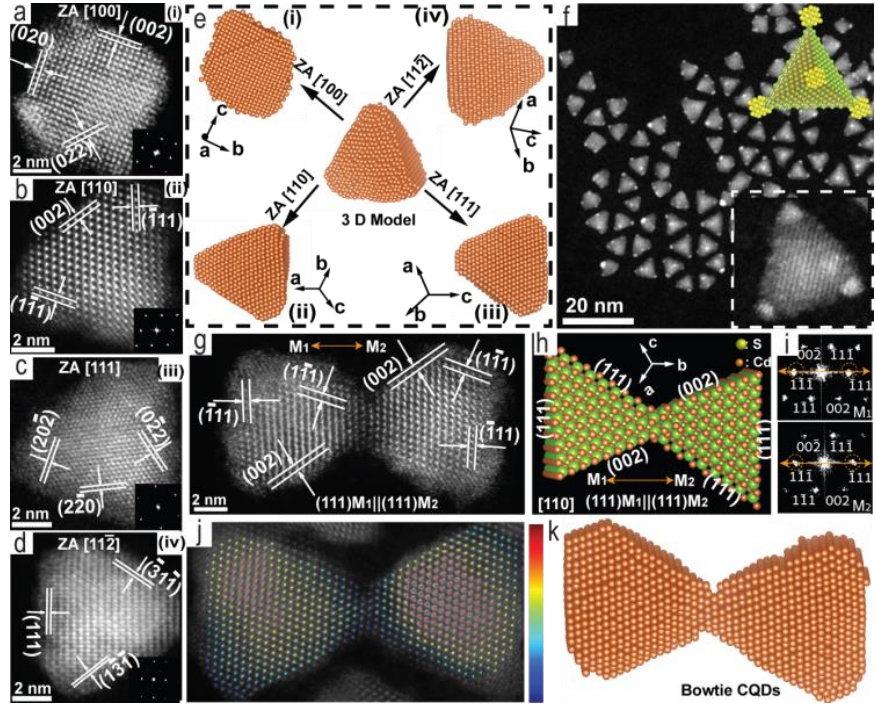

**Figure. 2.** Binding motifs in the bowtie structures: The raw STEM images and related FFT patterns of tetrahedral zinc-blende CdSe/CdS CQDs monomer viewed under different ZA of (a) [100], (b) [110], (c) [111], and (d) [11$\bar{2}$]. (e). The computed 3-D reconstruction model of tetrahedral ZB CdSe/CdS QD monomers viewed under ZA of [100], [110], [111], and [11$\bar{2}$], corresponding to frames a-d, respectively. (f). TEM images of CdSe/CdS-Au hybrid structures demonstrating the tip reactivity of the CQD monomers. The cartoon represents the CdSe/CdS-Au hybrid structure. Raw STEM image (g), atomic structure model (h), FFT patterns (i), and computed 3-D reconstruction model (i, k), of a fused bowtie nanoantenna.

An attribute essential for the bowtie formation, is the high reactivity of the tips favoring linker binding leading to tip-to-tip attachment. The preferential tip reactivity is directly demonstrated by the region selective gold domains formation upon synthesis of Au@CdSe/CdS hybrid structures [25] (Figure 2f). The gold deposition occurs selectively onto all four tips of the tetrahedral CQDs with nearly no growth on the facets (Figure S13). This high reactivity of the tip regions is expected considering that before reconstruction, ignoring carboxylate surface ligand coverage, atoms at the tetrahedron tips manifest three dangling orbitals, while on the outer facets, there are only one or two dangling orbitals per atom (Figure S12e). Even after surface reconstruction, the tips region remains a high energy site with preferential reactivity over the facet atoms. Therefore, the linkers preferentially bind to the exposed free tip of the CQDs that are bound to the silica template which is also less sterically hindered compared with other crystal sites. Upon adding the secondary CQD monomers, which also manifest higher tip reactivity, tip-to-tip linked dimers attached by the molecular linker are therefore preferentially formed.

Examining the attachment and orientation relationship in the bowtie structures after fusion by HRSTEM and model analysis, also supports the abovementioned formation mechanism. Such analysis is demonstrated in Fig. 2g-h. In this case, both the left and right monomers, $M_1$ and $M_2$, respectively, manifest in the HRSTEM image (002) and ($\bar{1}$11) lattice planes (Figure S12), typical of the zinc-blende structure observed under the [110] zone axis. We conclude that in this case there is an homonymous (111) planes connection formed between the two monomers defined as a

(111)$_{M1}$||(111)$_{M2}$ binding relationship for this type of bow-tie structure, as is well reproduced by the computed 3D reconstruction model. Another example for such homonymous (111) planes connection, viewed along a different zone axis, is shown in Figure S14.

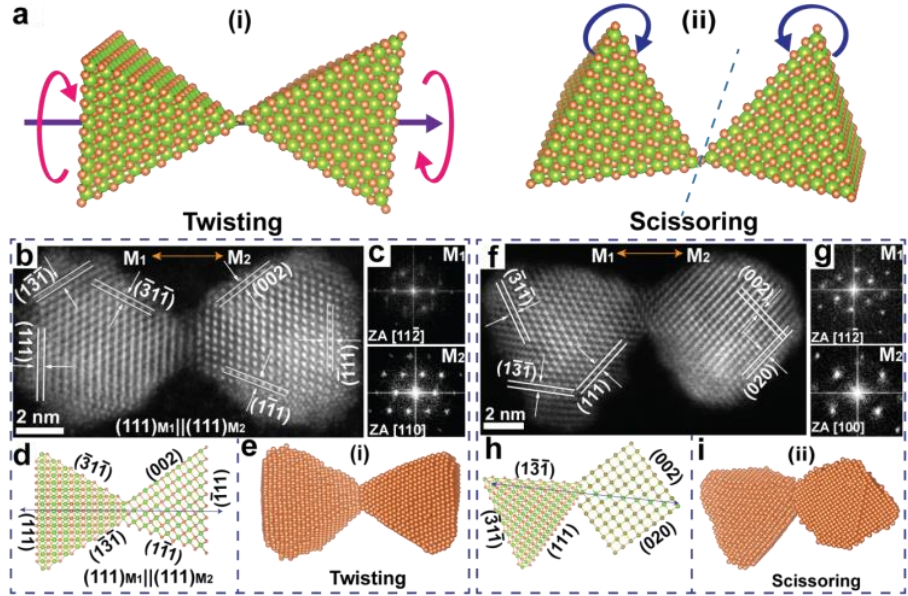

**Figure. 3.** Conformations of bowtie CQDs from twisting and scissoring rotations of the monomers. Two types of rotational motion, twisting and scissoring, are possible prior to the fusion of the tips forming the bowtie structures as indicated in (a), with the respective rotation axis indicated by the blue arrows. STEM images (b, f), the related FFT patterns (c, g), the related 2-D atomic structures (d, h), and computed 3D reconstruction structural models (e, i) for bowtie CQDs demonstrating twisting and scissoring rotational degrees of freedom, respectively.

Further modifications of the detailed structure of the bowtie were also observed, related to the variations in relative monomer orientations prior to and during the fusion reaction, assigned to some rotational freedom under the constrained crystal symmetry. We found that two types of rotations, twisting and scissoring modes of the monomer crystals along the bowtie axis, can explain the structural diversity (Figure 3, Figure S15). Figure 3b-e presents a case of a bowtie with twisted orientation, while still manifesting the (111)$_{M1}$||(111)$_{M2}$ binding relationship. Through the analysis and 3-D reconstruction, we extract a twist angle of ±30º and 90º between the ZA of [110] and [11$\bar{2}$]. Possible twist angles are in general constrained by the *fcc* crystal structure. Figure 3j-m manifests a bowtie with a scissoring rotation. Combination of both twisting and scissoring is also possible (Figure S15e-h). We observed relatively high abundance of bowtie CQDs structures with the twisted attachment motif compared with the scissoring motif, consistent with less interfacial strain upon twisting.[26,27] In case of facet attachment rather than tip attachment, as seen for example in the wurtzite dimers, the attachment is more constrained, and hence the fused interface planes along the connection axis are immobilized with limited rotation. In the present case of zinc-blende tetrahedral CQDs tip-to-tip attachment forming the bow-tie nanoantenna, the fused interfacial area is reduced significantly and the twisting and scissoring rotational modes can be accommodated upon fusion.

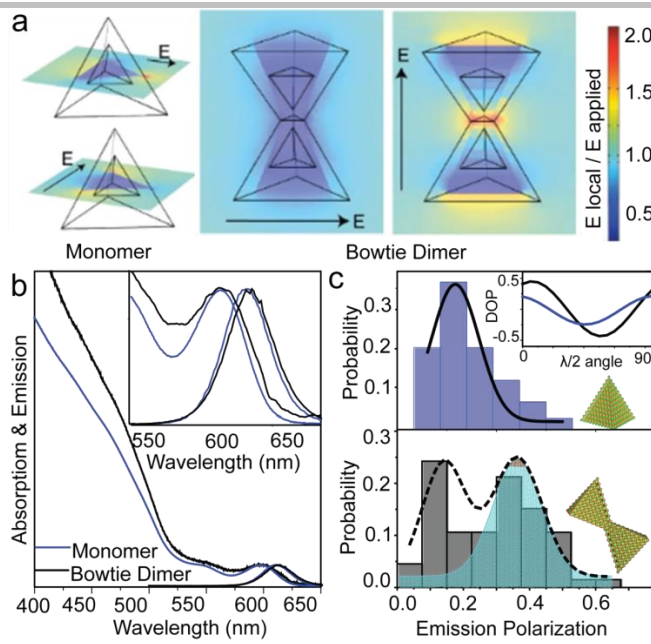

**Figure. 4.** Bowtie nanoantenna features of preferred polarization and Electric Field Hotspot. (a). Calculated electric field distribution in a tetrahedral monomer CQD and a bowtie dimer. The electric fields outside the tips of the tetrahedral monomers are enhanced to small extent symmetrically with no particular directional preference with the two components of the electric field depolarized to the same extent. In contrast, the component of the electric field along the long axis of the dimer is enhanced, while the perpendicular electric field is decreased to around half of the external field, leading to preferential polarization along the bowtie axis. (b) The absorption and emission spectra for the monomer (blue) and bowtie dimer (black). (c) Measured single particle emission polarization characteristics of the bowtie dimers compared to the monomers. Inset represents single-particle polarization dependence measurements as a function of analyzing polarizer angle showing the polarization, $(I_H - I_V)/(I_H + I_V)$, fitted to a sinusoidal function (see the supporting information) for a typical monomer (blue), and dimer (black). The histograms of the extracted degree of polarization from single bowtie dimers (black) and monomers (blue) clearly demonstrate the enhancement in the dimeric structure. The bimodal statistics in the bowtie dimer sample are associated with a fraction of monomers. Deconvolution using the tetrahedral monomer histogram parameters allows extraction of the polarization from the bowtie nanoantenna selectively.

After establishing the successful synthesis and characterization of the CQD bowtie structure, we discuss its merit under the classification of a nanoantenna. Bowtie nanoantennas are characterized by directionality of the electromagnetic field along the bowtie axis and its concentration forming a hot spot at the epicenter. We next demonstrate computationally and experimentally, the expression of this in the CQD bowtie structure. Figure 4a shows a cross section of the calculated distribution of the internal electric field in response to two perpendicular polarizations of the applied field, comparing between the CQD tetrahedral monomers and the bowtie dimers. The calculation, performed using Comsol Multi-Physics to solve the Poisson equation, utilizes the dielectric constants, $\varepsilon_\infty$, of both CdSe (6.2) and CdS (5.5) at the core and shell regions, [28] respectively, [29] with $\varepsilon_\infty = 1$ in the surroundings (SI for further details, Figure S16-17). The tetrahedral monomer, characterized by high $C_{3V}$ symmetry, does not manifest directionality of the electric field response. It shows the expected electric field enhancement at the sharp edges and the apex region outside the surface (low dielectric constant surrounding), with reduced electric field inside the nanocrystal (high dielectric constant material). In contrast, the bowtie structure has a clear directional response. Upon applying the field perpendicular to the dimer axis, the internal electric field is reduced by more than half. Applying a field parallel to the dimer axis, leads to a doubled enhanced internal field at the bowtie epicenter, inside the dimer, creating a well localized hotspot. Also away from the epicenter, the internal field is significantly higher as compared to the case of the perpendicular polarization. Notably, the bowtie antenna shape matters here, as similar calculation on spherical CQD dimers did not show any internal hotspot formation (Figure S18).

The emission polarization, $P = [(I_H - I_V)/(I_H + I_V)]$, where $I_H$, $I_V$ are the horizontal and vertical intensities, respectively, is a first indicator to test the CQD nanoantenna functionality.[30] In order to estimate the expected *P* of both the monomers and dimers we calculate for the structures the electron wavefunction, $\Psi_e(r)$, and the hole wavefunction, $\Psi_h(r)$, using an effective mass based model. We note in passing that for these CQD bowtie dimers, only small hybridization energies are expected in light of the relatively large core/shells and the constrained epicenter in-between the monomers presenting a tunneling barrier. We use the following expression for *P*:

$$P = \frac{\left\langle |E_{II}|^2 * \Psi_e(r) * \Psi_h(r) \right\rangle - \left\langle |E_\perp|^2 * \Psi_e(r) * \Psi_h(r) \right\rangle}{\left\langle |E_{II}|^2 * \Psi_e(r) * \Psi_h(r) \right\rangle + \left\langle |E_\perp|^2 * \Psi_e(r) * \Psi_h(r) \right\rangle} \qquad \text{eq. 1}$$

Equation 1 represents the relative emission intensity in the two orthogonal orientations by taking the volume average (indicated by the brackets), of the square of the two electric field components $E_{II}$ and $E_\perp$ with a weight function related to the wavefunctions, $\Psi_e(r) * \Psi_h(r)$ (SI for details). For the monomer in the particular orientation of Figure 4c the emission is found to be nearly unpolarized, $P \sim 0.05$, while for the bowtie dimer the emission is polarized with $P = 0.42$. We note that this is a high degree of polarization considering that its sole source is the dielectric antenna effect.

Experimentally. the degree of polarization, *P*, was measured by single particle emission polarization measurements comparing the tetrahedral monomers with the bowtie dimers (See SI and Supplementary Figures 19-20 for experimental details). Figure 4c presents the polarization degree histograms for both samples. Note that the dimer sample consists of ~30% of the unreacted monomers, as revealed by TEM measurements, which explains the bimodal feature in its polarization histogram. While the tetrahedral monomers manifest a low mean polarization degree of 0.15, significantly higher polarization is observed for the bowtie dimers reaching up to 0.5, with a mean value of 0.38 (excluding the monomer fraction), well correlated with the calculated polarization value. The distribution in the polarization values can be attributed to the different tilt angles of the nanocrystals with respect to the substrate as verified by calculating the effect of different orientations with respect to the objective lens on the emission polarization values (Figure S16-18). Additional sources of the differences between the experimental and calculated emission polarizations can be attributed to deviation of the position of the cores from the center. The highest emission polarization is obtained for a flat lying dimer with its bowtie axis parallel to the substrate plane, which is also a plausible geometry in the actual single particle measurements. Considering these variations, the experimental values are in good agreement with the simulations, demonstrating the directionality of the CQD bowtie nanoantenna.

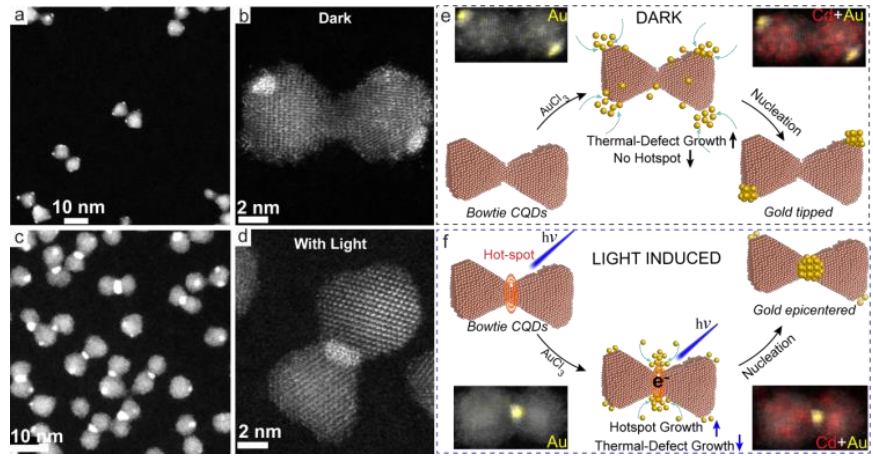

**Figure. 5.** Nanoantenna driven photochemical reaction of gold growth. TEM images following the growth of gold domains on bowtie CQD nanoantennas in the dark (a-b) and in presence of light irradiation (c-d). In the dark conditions, the gold domain growth is restricted to the reactive tips, whereas the light irradiation drives selective metal domain growth at the hotspot in the bowtie epicenter. The suggested gold growth mechanisms on the bowtie CQDs nanoantenna in dark (e) and light (f) induced conditions with corresponding representative elemental-STEM overlaid mapping images in the insets. The light illumination reaction was performed with a UV lamp (365 nm ~20mW/cm$^2$) for 30 min.

Beyond directionality, we discovered that the CQD scaled down bowtie nanoantenna bears a unique feature of concentrated photocatalytic activity at its epicenter, which manifests a clear electric field hot spot in the calculations (Figure 4a). We demonstrate this by performing a photoinduced chemical reaction of gold growth. As discussed above, in dark conditions selective gold domain formation is observed at the highly reactive tips of the tetrahedral monomers. Similarly, under dark conditions, gold domain growth on the free tips of the CQD bowtie is observed. But carrying the reaction at low temperature of 4 °C suppresses this thermal growth route and yields only small tips (Figure 5a-b, Figure S21-22). Upon UV (365nm) light irradiation during the growth with similar Au concentration and temperature, there is still also some growth on the free tips in some cases but this is significantly suppressed. Instead, formation of a gold domain at the bowtie epicenter is clearly observed (Figure 5c-e, Figure S23-24). The selective gold deposition at the epicenter under illumination correlates with the presence of the electric field hotspot, while the thermal induced gold growth at the tips is restricted substantially by reducing the reaction temperature. [23,31,32] Indeed, at room temperature under illumination, enhanced growth on the free tips is resolved, on account of epicenter growth (Figure S 25). The hotspot leads to higher concentration of photo-generated carriers at the bowtie epicenter, which is also a region of potential trap sites leading to accumulation of electrons for the selective light induced gold reduction at this position. Once gold clusters are formed, the photocatalytic action further leads to additional gold deposition preferentially at the epicenter. The light induced growth at low temperature effectively competes with thermal growth suppressing the gold deposition at the bowtie tips, and the nanoantenna effect dominates, where the electric field hotspot acts as a photochemical reaction center (Figure 4f, Figure S 26).

## Conclusion

Metallic bowtie nanoantenna structures are fabricated typically by e-beam lithography methods and manifest electric field enhancement factors as high as 1000, but beyond the fabrication and scale-up challenges, the functionality in emission enhancement suffers from competing quenching by the metallic structures.[5,7,33,34] Alternatively, dielectric Silicon nanoantennas function to control the optical transitions at the telecom wavelengths and a moderate modulation of radiative rate for quantum emitters at the epicenter has been achieved with the metal-dielectric hybrid antenna. [3,35,36] The bottom-up assembled scaled down semiconductor bowtie nanoantenna on the other hand, provides moderate field enhancement that can be further increased by larger CQD dimensions and using higher dielectric semiconductors. This may lead to a stand-alone single photon emitter within the bottom-up prepared bowtie nanoantenna. Notably, the bowtie nanoantenna manifests unique functionality as a photochemical reaction antenna utilizing the electric field and charge concentration at the bowtie epicenter, with relevance towards energy conversion and additional photocatalytic applications.

## Acknowledgements

The research leading to these results has received financial support from the European Research Council (ERC) under the European Union's Horizon 2020 research and innovation programme (grant agreement No [741767]). J.B.C and S.K acknowledge the support from the Planning and Budgeting Committee of the higher board of education in Israel through a fellowship. U.B. thanks the Alfred & Erica Larisch memorial chair. Y.E.P. acknowledges support by the Ministry of Science and Technology & the National Foundation for Applied and Engineering Sciences, Israel. We thank Dr. Inna Popov for helpful discussions.